# Sudden drop of fractal dimension of electromagnetic emissions recorded prior to significant earthquake

S. M. Potirakis[a], G. Minadakis[b], and K. Eftaxias[c]

a. Department of Electronics Engineering, Technological Education Institute (TEI) of Piraeus, 250 Thivon & P. Ralli, GR-12244, Aigaleo, Athens, Greece, spoti@teipir.gr .
b. Department of Electronic and Computer Engineering, Brunel University Uxbridge, Middlesex, UB8 3PH, U.K., george.minadakis@brunel.ac.uk
c. Department of Physics, Section of Solid State Physics, University of Athens, Panepistimiopolis, GR-15784, Zografos, Athens, Greece, ceftax@phys.uoa.gr .

**Abstract**

The variation of fractal dimension and entropy during a damage evolution process, especially approaching critical failure, has been recently investigated. A sudden drop of fractal dimension has been proposed as a quantitative indicator of damage localization or a likely precursor of an impending catastrophic failure. In this contribution, electromagnetic emissions recorded prior to significant earthquake are analysed to investigate whether they also present such sudden fractal dimension and entropy drops as the main catastrophic event is approaching. The pre-earthquake electromagnetic time series analysis results reveal a good agreement to the theoretically expected ones indicating that the critical fracture is approaching.

*Keywords*: fractal dimension; electromagnetic precursors; large scale fracture; earthquakes

**1. Introduction**

Over the past two decades, considerable effort has been devoted to the study of damage and fracture in disordered media (e.g., rocks). Recently, Lu et al. (2005) proposed the detection of a sudden drop of fractal dimension (FD) as a precursor for an impending catastrophic failure in disordered media. They found that the FD or entropy of the spatial distribution of microcracks/voids decreases as the damage in a disordered media evolves, and that a sudden-drop of FD provides a quantitative measure of the damage localization (or the clustering degree of microcracks/voids), which might be viewed as a likely precursor prior to a final catastrophic failure.

Earthquakes (EQs) are large-scale fracture phenomena in the Earth's heterogeneous crust. The Earth's crust is clearly extremely complex. However, despite its complexity, there are several universally holding scaling relations. In particular, the aspect of self-affine nature of faulting and fracture is widely documented based on analyses of data from both field observations and laboratory experiments, and studies of failure precursors on the small (laboratory) and large (EQ) scale. Indeed, a number of studies indicate a similarity of the statistical properties of the laboratory seismicity, in terms of acoustic and electromagnetic (EM) emission, on one hand, and seismicity at the geological scale, in terms of energy, on the other hand. It is found that the distribution of acoustic emission (AE) / electromagnetic emission (EME) pulses over energy obeys a power law, namely, corresponds to the Gutenberg-Richter frequency-magnitude relation. Other studies found that AE has a spatial fractal structure typical of seismicity as well. Focal mechanism solutions of AE are similar in type to those of EQ foci. Studies of time sequences of AE / EME pulses reveal effects typical of EQ sequences. The high degree of similarity between EQs and fractures strongly supports the fracture hypothesis of EQs. It is a risky practice to extend findings rooted in laboratory experiments to a geophysical scale. However, one cannot ignore the comparison between failure precursors at laboratory and geophysical scales. Notice, it has been early suggested that, the mechanism of the shallow EQs is apparently some sort of laboratory fracture process (Mogi 1962; Scholz 1968).

Fracture induced fields allow a real-time monitoring of damage evolution in materials during mechanical loading. EME in a wide frequency spectrum ranging from kHz to MHz are produced by cracks' opening, which can be considered as the so-called precursors of general fracture (Papadimitriou et al. 2008 and references therein). These precursors are detectable both at laboratory (Rabinovitch et al. 2001; Bahat et al. 2005) and geophysical scale (Gokhberg et al. 1995; Hayakawa and Molchanov 2002; Uyeda et al. 2009). Our main observational tool is the monitoring of the fractures which occur in the focal area before the final break-up by recording their kHz-MHz EME. Recent results indicate that these pre-seismic EM time series contain information characteristic of an ensuing seismic event (Papadimitriou et al. 2008; Kapiris et al. 2004; Contoyiannis et al. 2005; Karamanos et al. 2006; Eftaxias et al. 2009). An improved understanding of the EM precursors, especially its physical basis, has direct implications for the study of EQ generation processes and EQ prediction research.

An important challenge in this field of research is to identify an emerged anomaly in a recorded EM time series as a pre-seismic one and correspond this to an individual stage of EQ preparation process. This paper focuses on this direction based on FD and entropy changes. An anomaly is defined as a deviation from normal (background) behaviour. Experimental and theoretical evidence indicates that as the final failure in the disordered medium approaches, the underlying complexity manifests itself in linkages between space and time, generally producing patterns on many scales and the emergence of a fractal structure which is analogous to a non-equilibrium phase transition (Eftaxias et al. 2004, Eftaxias et al. 2006, and references therein). In this situation the FD and related exponents can be used as a quantitative measure of the character of the system (Olemskoy and Flat 1993). In order to develop a quantitative identification of EM precursors it is checked whether the findings of Lu et al. (2005) can be confirmed on EM time series recorded prior to significant EQs. It is also investigated whether the findings of Lu et al. (2005) are detected in different energy scales, namely in laboratory scale, in fault scale and in regional seismicity scale. The analysed EM precursors are related to fault scale phenomena (fault activation), and are expected to behave as a magnified image of the laboratory seismicity and a reduced image of the regional seismicity. It is noted that, very recently the seismicity around the activated fault and the corresponding EME precursor signals have been analyzed using a non-extensive (Tsallis statistics based) frequency-magnitude law (Minadakis et al., 2012b). It was found that both of them, prior to significant EQ, share common (increased) non-extensive $q-$parameter values. The increase of super-extensivity ($q > 1$) signifies further privileging of prominent events. The corresponding estimated $b-$values were found reduced. This finding renders the anticipated agreement to the findings of Lu et al. (2005) quite reasonable, since FD is proportional to $b-$value (Lu et al. 2005).

## 2. Fractal dimension

The term Fractal Dimension (FD) generally refers to any of the dimension used for fractal characterization. In fractal geometry, the FD is a statistical quantity that gives an indication of how completely a fractal appears to fill the space, as one zooms down to finer and finer scales, accordingly there are many specific definitions of FD. The FD is a measure of how complicated a self-similar figure or time series is. According to Mandelbrot (Mandelbrot 1983), a fractal is a set for which the Hausdorff-Besicovitch dimension ($D_h$) strictly exceeds the topological dimension.

Hence, every set with a non-integer dimension $D$ is a fractal.

Various algorithms for calculating FD have been developed. Meanwhile, a general solution is not available. It is often said, e.g., (Schmittbuhl et al. 1995), that at least two different algorithms are needed for a faithful representation of the FD of time series. Here, the box-counting algorithm and the Higuchi's algorithm are adopted to estimate the FD of pre-EQ EM time series.

### 2.1 Box-counting algorithm

The box-counting algorithm estimates the space filling properties of a curve. In this approach, the curve is covered with a grid of area elements ("boxes") of the same size, then the number of elements of a given size which are necessary to completely cover the curve are counted as the element size successively reduces. As the size of the area element approaches zero, the total area covered by the area elements will converge to the area covered by the curve. If the curve is fractal, the following relation holds:

$$M(r) \underset{r \to 0}{\sim} (1/r)^{D_B} \qquad (1)$$

or, as usually expressed:

$$D_B = \lim_{r \to 0} \frac{\log M(r)}{\log(1/r)} \qquad (2)$$

where $M(r)$ is the total number of boxes of size $r$, required to cover the curve entirely. The regression slope, $D_B$, of the straight line formed by plotting $\log M(r)$ against $\log(1/r)$ indicates the degree of complexity, or FD.

The above described procedure, also called "grid method", is usually applied to time series either by employing two-dimensional processing, which is highly computationally demanding, e.g.,. (Raghavendra and Narayana Dutt 2010), or by employing appropriately chosen thresholds, e.g., (Zhou et al. 2006). Here, a computationally efficient variation which is described in detail in (Shoupeng et al. 2007) was used.

**2.2 Higuchi's algorithm**

Higuchi's algorithm (Higuchi 1988) for FD estimation is based on curve length measurement. The algorithm estimates the mean length of the curve, by using a segment of $k$ samples as a unit of measure. In the case of a time series $\{x_n\}$, $n = 1, 2, ..., N$, where $x_n = x(t_n)$, with $t_n = nT$, $n = 1, 2, ..., N$, and $T$ being the sampling period, one can define a set of $k$ new time series, $\{x_k^m\}$, $m = 1, 2, ..., k$, defined as:

$$x_k^m = \left\{ x(t_m), x(t_{m+k}), x(t_{m+2k}), ..., x\left(t_{m+\text{int}[(N-m)/k] \cdot k}\right) \right\} \qquad (3)$$

where $\text{int}[(N-m)/k]$ denotes the integer part of $(N-m)/k$, $m$ and $k$ are integers defining the initial time value and the time interval between successive considered values, respectively. The length of each new time series can be defined as follows:

$$L_m(k) = \frac{N-1}{\text{int}[(N-m)/k] \cdot k^2} \sum_{i=1}^{\text{int}[(N-m)/k]} \left| x(t_{m+ik}) - x(t_{m+(i-1)k}) \right| \qquad (4)$$

The term $N - 1/\text{int}[(N-m)/k]k$ represents the normalization factor for the curve length of subset time series $x_k^m$. Calculating the mean length of the curve for each $k$, $\langle L(k) \rangle$, as $\langle L(k) \rangle = k^{-1} \sum_{m=1}^{k} L_m(k)$, a set of $(k, \langle L(k) \rangle)$ pairs can be formed for $k = 1, 2, ..., k_{\max}$.

If the curve is fractal, the following relation holds:

$$\langle L(k) \rangle \propto (1/k)^{D_H} \qquad (5)$$

The regression slope, $D_H$, of the straight line formed by plotting $\log \langle L(k) \rangle$ against $\log(1/k)$ is the Higuchi FD.

**3. Tsallis entropy in its symbolic form**

Symbolic time series analysis is a useful tool for modeling and characterization of nonlinear dynamical systems/signals (Voss et al., 1996). It is a way of coarse-graining or simplifying the description (Hao, 1989).

In the framework of symbolic dynamics, time series are transformed into a series of symbols by using an appropriate partition which results in relatively few symbols. After symbolization, the next step is the construction of sequences of symbols ("words" in the language of symbolic dynamics) from the series of symbols by collecting groups of symbols together in temporal order.

We restrict ourselves to the simplest possible coarse-graining of the time-series under study. This is given by choosing a threshold $C$ (the mean value is used here as a threshold, however note that different values lead to similar results (Karamanos et al. 2005)) and assigning the symbols "1" and "0" to the time-series, depending on whether it is above or below the threshold (binary partition). In this way, each time window of the original time-series for a given threshold is transformed (it is encoded) into symbolic sequences, which contains "linguistic" or "symbolic dynamics'" characteristics. Thus, we generate a symbolic time series from a 2-letter ($\lambda = 2$) alphabet (0, 1), e.g., 0110100110010110 . . . . Reading the sequence by words of length $L = 2$ one obtains 01/10/10/01/10/01/01/10/ . . . . The number of all possible kind of blocks is 00, 01, 10, and 11. Thus, the required probabilities for the estimation of the Tsallis entropy, $p_{00}$, $p_{01}$, $p_{10}$, $p_{11}$ are the fractions of the blocks 00, 01, 10, 11 in the symbolic time series, namely, 0, 4/16, 4/16, and 0, correspondingly in this example.

In a symbolic time series of $W$ symbols, $\{A_i\}$, $i = 1, 2, ..., W$, one can read it by words of length $L = m$, $(m < W)$. For each word length, there are $\lambda^m$ possible combinations of the symbols that may be found in a word, here $\lambda^m = 2^m$, since $\lambda = 2$. The probability of occurrence $p_j^{(m)}$ of the $j-$th combination of symbols ($j = 1, 2, ..., 2^m$) in a word of length $m$, can be denoted as:

$$p_j^{(m)} = \frac{\text{\# of the } j-th \text{ combination found in words of length } m}{\text{total \# of words of length } m \text{ (by lumping)}}. \qquad (6)$$

Tsallis entropy has been examined in terms of symbolic dynamics, yielding interesting results, e.g., (Papadimitriou et al. 2008; Kalimeri et al. 2008; Potirakis et al. 2012). In terms of symbolic dynamics, the Tsallis entropy for word length $m$, $S_q(m)$, is:

$$S_q(m) = k \frac{1}{q-1} \left(1 - \sum_{j=1}^{2^m} \left[p_j^{(m)}\right]^q \right). \qquad (7)$$

The value of $q$ is a measure of the non-extensivity of the system. The cases $q > 1$ and $q < 1$, correspond to sub-extensivity, or super-extensivity, respectively. We may think of $q$ as a bias-parameter: $q < 1$ privileges rare events, while $q > 1$ privileges prominent events. We emphasize that *the parameter $q$ itself is not a measure of the complexity of the system but measures the degree of non-extensivity of the system*. Broad symbol-sequence frequency distributions produce high entropy values, indicating a low degree of organization. Conversely, when certain sequences exhibits high frequencies, low values are produced, indicating a high degree of organization.

Concerning the symbolic analysis, we note the following particularity: the results are originally derived for a number of word lengths. This means that for every word length a time-series of entropy is produced rendering the results three-dimensional, with third dimension the parameter of word length. In some cases, it is considered sufficient to present the results for a specific word length, e.g., (Papadimitriou et al. 2008; Kalimeri et al. 2008). However, for the sake of generality a dimension reduction method has been recently proposed taking into account the results for all word lengths yet leading to two-dimensional symbolic analysis results (Potirakis et al. 2012) and this is applied here.

## 4. Electromagnetic data analysis

A way to examine transient phenomena in a pre-seismic EM time series is to analyze it into a sequence of distinct time windows. The aim is to discover a clear difference of fractal characteristic and entropy as the catastrophic event approaches.

The well documented pre-seismic EM signal associated with the Athens EQ, with magnitude 5.9, occurred on 7 September 1999, e.g., (Kapiris et al. 2004; Contoyiannis et al. 2005; Karamanos et al. 2006), is employed as a test case. The analysed fragment of the candidate precursor time series is shown in Fig. 1a. It corresponds to an 11 days period from 28 August 1999, 00:00:00 (UT), to 7 September 1999, 23:59:59 (UT), of the EM time series associated with the Athens EQ. The EM time series recorded on the 10 kHz band associated with the Athens EQ has been analysed. Although the analysis of the time series associated to Athens EQ is the only which is presented here, the analysis of pre-seismic EM time series associated to other EQs has been verified to lead to similar results.

Figure 1b depicts the temporal evolution of the symbolic Tsallis entropy (with non-extensive parameter $q = 1.8$ (Papadimitriou et al. 2008)) for time windows of 3020 samples length, with 50% overlapping, and running time average of two windows with 50% overlapping (Potirakis et al. 2012). It is noted that for the entropy calculation only the time windows satisfying a good stationarity requirement, checked by a relevant stationarity criterion (Karamanos et al. 2005), were considered. This is the reason why a sliding window has been adopted in this case. Non-overlapping windows would exclude parts of the time-series as non-stationary, enough to result to an incomplete picture for entropy evolution.

Figures 1c and 1d show the evolution with time of the Higuchi and the box-counting FD, respectively. The FD analysis has been performed on successive (non-overlapping) time windows of 3020 samples size. It must be noted here that all practical FD estimates are very sensitive to numerical or experimental noise, e.g., (Raghavendra and Narayana Dutt 2010). Overlapping windows were not necessary in this case since stationarity is not required before the FD estimation and thus all windows are analysed. The presence of noise leads to FD estimates which are higher than the actual FD. This effect, although intuitively expected since noise is increasing time series roughness, can easily be verified by employing known fractal time series (e.g., fractional Brownian motion, fBm, time series of known roughness). The analysed field-acquired EM time series is certainly contaminated by measurement noise. The different sensitivity to the measurement noise leads to the observed differences between the FD estimates through the two employed algorithms. Nonetheless, the important objective in the context of this study is to detect the temporal changes in FD, not its exact value.

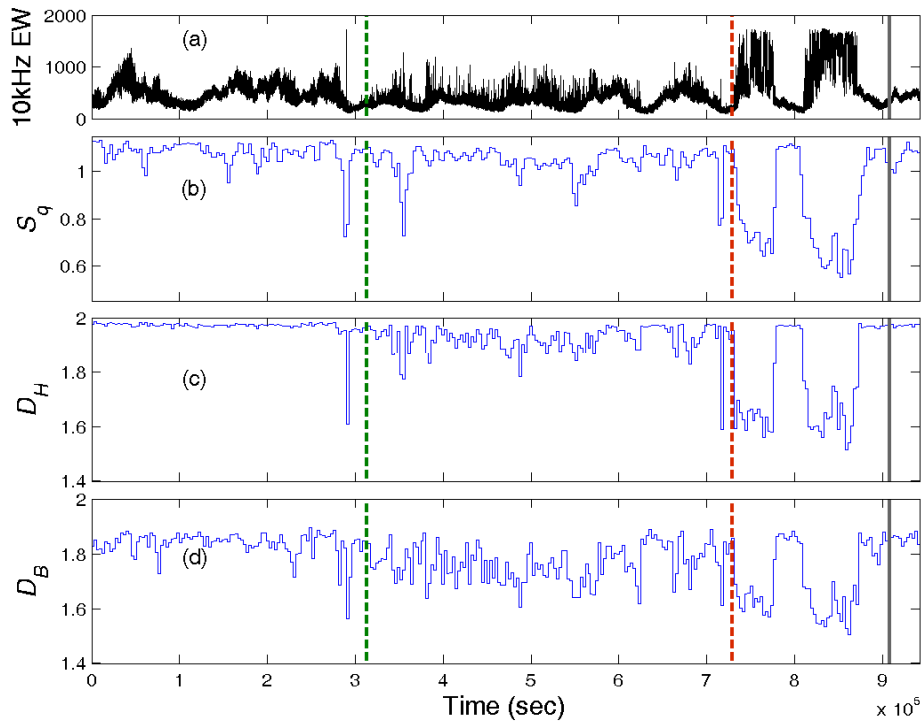

**Fig.1** (a) Part of the recorded time series of the 10 kHz (East–West) magnetic field strength (in arbitrary units) covering an 11 days period from 28 August 1999, 00:00:00 (UT), to 7 September 1999, 23:59:59 (UT), associated with the Athens EQ.

(b) Symbolic Tsallis entropy, $S_q$, of the specific time series fragment. (c) Higuchi FD, $D_H$, of the specific time series fragment. (d) Box-counting FD, $D_B$, of the specific time series fragment. The common horizontal axis is the time (in s)., denoting the relative time position from the beginning of the analysed part of the EM recording. The (left) vertical broken green line roughly indicates the start of the candidate precursor. The (right) vertical broken red line indicates where the damage evolution of the fault approaches the critical point. The vertical solid grey line indicates the time of the Athens EQ occurrence.

The entropy evolution, shown in Fig. 1b, suggests the presence of two distinct phases of the underlying fracture process, as has already pointed out (Papadimitriou et al. 2008; Kalimeri et al. 2008). An initial stage could be identified to begin around the time indicated by the (left) green vertical line which signifies the emergence of the candidate precursor after a long background noise period (here only a short part of this background noise period is presented, for the complete signal please refer to Fig. 1 of (Papadimitriou et al. 2008)). Lower entropy values, compared to that of the background noise, can be observed between the green (left) and the red (right) broken vertical lines, although sparsely distributed in time. On the other hand, the entropy values suddenly drop during the two strong EM pulses signifying a different behaviour, a new distinct phase in the tail of the EQ preparation process which is characterized by a significantly higher degree of organization and lower complexity in comparison to that of the preceding phase.

Correspondingly, a largely stable high FD until the (left) green broken vertical line is observed on figures 1c and 1d. It is suggested that the time position of the (left) green broken line roughly indicates the start of the candidate precursor. From the (left) green broken vertical line to the (right) red broken vertical line there is a continuous alternation of low and high FD. This feature probably corresponds to a phase of small scale distinct fracture events which spatial distribution decreases as the damage evolves. Very recently, this phase has been corresponded to the fragment of smaller entities existent within the fault area (Minadakis et al., 2012a). The whole fault system however seems to preserve a stable state during this specific phase. Finally, after the (right) red broken vertical line, two sudden drops of FD signify that the damage evolution of the fault approaches the critical point, when the strong teeth (asperities) sustaining the fault are fractured (Minadakis et al., 2012a).

This final phase of precursory EM phenomenon corresponds to the sudden drop of FD and entropy, which was reported by Lu et al. (2005) as a quantitative measure of the damage localization (or the clustering degree of microcracks/voids), and a likely precursor prior to a final catastrophic failure.

Following the analysis presented by Lu et al. (2005), it is noted that the sudden drop in FD corresponds to the part of the stress ($\sigma$)-strain ($\varepsilon$) curve ($\sigma = g(\varepsilon)$, please refer to Fig. 1 in (Lu et al. 2005)) where stress reaches its maximum ($\frac{\partial \sigma}{\partial \varepsilon} = 0$) and after. During this part of the stress-strain curve, the left part of equation (5) in (Lu et al. 2005), i.e., the term ($\frac{\partial \sigma}{\partial \varepsilon} - \frac{\sigma}{\varepsilon}$), takes suddenly a high negative value since $\frac{\partial \sigma}{\partial \varepsilon}$ is rapidly decreasing negative and $\frac{\sigma}{\varepsilon}$ remains positive (although decreasing). Since $\frac{\partial \sigma}{\partial \varepsilon} - \frac{\sigma}{\varepsilon} \propto \frac{\partial D_f}{\partial t}$ (please refer to equation (5) in (Lu et al. 2005)), this means that during the specific part of the curve the FD suddenly drops. Therefore, it is argued that these two sudden drops correspond to the fracture of the asperities which were sustaining each one of two different faults. The specific argument is further supported by various different experimental and analysis data, e.g., (Kontoes et al. 2000, Eftaxias et al. 2001), which indicate that the specific Athens EQ was related to two faults: the main fault segment was responsible for 80% of the total energy released, with the secondary fault segment for the remaining 20%. Moreover, a seismic data analysis indicates that there was probably a subsequent EQ with magnitude 5.3 after about 3.5 s of the main event (Eftaxias et al. 2001).

Finally, scale invariance is expected to be valid. The findings of Lu et al. (2005) which were confirmed at fault scale by analysing the associated EME are expected to be also detectable at different energy scales, namely at laboratory scale, and at regional seismicity scale. The analysed

EM precursors are related to fault scale phenomena (fault activation), and are expected to behave as a magnified image of the laboratory seismicity and a reduced image of the regional seismicity.

Indeed, during the deformation of rock in laboratory experiments, small cracking events emerge which radiate elastic waves in a manner similar to EQ (Scholz 1968; Ponomarev et al. 1997). These emissions were found to obey the "Gutenberg-Richter type" relation. AE from rock fracturing present a significant fall of the observed $b-$values as the main event approaches, i.e. indicate a significant decrease in the level of the observed $b-$values immediately before the critical point, e.g., (Mogi 1962; Scholz 1968; Ponomarev et al. 1997; Weeks et al. 1978; Lei and Satoh 2007; Li et al. 2009). The sudden reduction of the $b-$value before the EQ occurrence is also reported at regional seismicity scale by several researchers, e.g., (Lu et al. 2005; Enescu and Ito 2001; Tsukakoshi and Shimazaki 2008; Wu et al. 2008). Moreover, it is widely known that FD is directly proportional to the $b-$values (Lu et al. 2005). Very recently, it was shown that in terms of Tsallis non-extensivity, the decrease of the $b-$value (and correspondingly, the FD) can be put in relationship with the increase of the parameter $q$, which quantifies the degree of non-extensivity of a system (Telesca, 2010a). Such parameter significantly increases before the occurrence of strong earthquakes (Telesca, 2010b). The above results for the non-extensive parameter $q$, and the interrelated $b-$value have been very recently verified for the seismicity around the activate fault, yet, extended for the EME precursory signals recorded prior to significant EQ (Minadakis et al., 2012b). Therefore, a sudden reduction of FD is observed at all three scales (laboratory, fault, regional seismicity).

## 5. Conclusions

EM time series recorded in the kHz band prior to significant EQ has been analysed in terms of FD and entropy. The results prove a sudden reduction of both quantities prior to the final catastrophic event. These results are in agreement to the theoretic and experimental results recently presented by Lu et al. (2005). They imply that the pre-EQ EM time series can be considered as a precursor of an impending EQ, indicating the time of the damage localization along the fault direction and the associated final fracture.